\documentclass[12pt,manuscript]{aastex}

\begin{document}

\pagestyle{plain}
9/8/2006
\vspace{1in}
\begin{center}
\textbf{Visible and Near-Infrared Spectrophotometry of the Deep Impact Ejecta of Comet 9P/Tempel~1}\\
\vspace{0.5in}
Klaus W. Hodapp$^{a,*}$, 
Greg Aldering$^b$,
Karen J. Meech$^c$,
Anita L. Cochran$^d$,
Pierre Antilogus$^h$,
Emmanuel P\'{e}contal$^i$,
William Chickering$^b$,
Nathalie Blanc$^j$,
Yannick Copin$^j$,
David K. Lynch$^e$,
Richard J. Rudy$^e$,
S. Mazuk$^e$,
Catherine C. Venturini$^e$,
Richard C. Puetter$^f$,
and Raleigh B. Perry$^g$\\

\vspace{0.5in}
$^a$Institute for Astronomy, University of Hawaii, 640 N. Aohoku Place, Hilo, HI 96720\\
$^*$Corresponding Author E-mail address: hodapp@ifa.hawaii.edu\\
\vspace{0.3in}
$^b$Lawrence Berkeley Lab, Physics Div., MS-50/232,
One Cyclotron Rd., Berkeley, CA 94720\\
\vspace{0.3in}
$^c$Institute for Astronomy, University of Hawaii,
2680 Woodlawn Drive, Honolulu, HI 96822\\
\vspace{0.3in}
$^d$McDonald Observatory, University of Texas at Austin,
1 University Station C1402, Austin, TX 78712-0259\\
\vspace{0.3in}
$^e$The Aerospace Corporation,
P.O. Box 92957, MS 266,
Los Angeles, CA 90009\\
\vspace{0.3in}
$^f$University of California, San Diego,
CASS 0424, 9500 Gilman Dr., La Jolla, CA 92093
San Diego, CA\\
\vspace{0.3in}
$^g$NASA Langley Research Center,
MS 160, Science Support Office,
Hampton, VA 23681\\
$^h$Laboratoire de Physique Nucl\'{e}aire et des Hautes Energies IN2P3 - CNRS -
Universit\'{e}s Paris VI et Paris VII, 4 Place Jussieu Tour 33 - Rez des chauss\'{e}e 75252 Paris Cedex 05\\
$^i$Centre de Recherche Astronomique de Lyon, 9, av. Charles Andr\'{e},\\
69561 Saint Genis Laval Cedex\\
$^j$Institut de Physique Nucle\'{e}aire de Lyon, UMR5822, CNRS-IN2P3, Universit\'{e} Claude Bernard Lyon 1,
F-69622 Villeurbanne, France

\end{center}

\vspace{0.5in}
\noindent
Pages:  28\\
Tables:  1\\
Figures: 8\\

\newpage

\noindent
Proposed Running Head: Spectro-Photometry of Deep Impact

\vspace{1.0in}

\noindent
Editorial correspondence to:\\
\vspace{0.3in}
Dr. Klaus W. Hodapp\\
Institute for Astronomy\\
640 N. Aohoku Place\\
Hilo, HI 96720\\
Phone: 808-932-2313\\
Fax:   808-933-0737\\
Email: hodapp@ifa.hawaii.edu\\

\newpage

\textbf{ABSTRACT}\\

We have obtained optical spectrophotometry
of the evolution of comet 9P/Tempel 1 after the
impact of the Deep Impact probe,
using the Supernova Integral Field Spectrograph (SNIFS) at
the UH 2.2m telescope, as well as simultaneous optical and infrared
spectra using the Lick Visible-to-Near-Infrared Imaging Spectrograph (VNIRIS) spectrograph. 
The spatial distribution and temporal evolution
of the ``violet band'' CN~(0-0) emission and of the 630~nm~[OI] emission 
was studied.
We found that CN emission centered on the nucleus increased
in the two hours after impact, but that this CN emission was delayed compared to the
light curve of dust-scattered sunlight. The CN emission also
expanded faster than the cloud of scattering dust.
The emission of [OI] at 630 nm rose
similarly to the scattered light, but then remained nearly constant
for several hours after impact.
On the day following the impact, both CN and [OI] emission concentrated 
on the comet nucleus had returned nearly to pre-impact levels.
We have also searched for 
differences in the scattering properties of the dust ejected
by the impact compared to the dust released under normal conditions.
Compared to the pre-impact state
of the comet, we find
evidence that the color of the comet was slightly bluer 
during the post-impact rise in brightness.
Long after the impact, in the following nights, the comet colors
returned to their pre-impact values.
This can be explained by postulating a change to a smaller particle
size distribution in the ejecta cloud, in agreement with the findings
from mid-infrared observatons, or by postulating a large fraction
of clean ice particles, or by a combination of these two.


\newpage

\section{Introduction}
As part of a coordinated simultaneous Earth-based 
observing campaign (Meech et al. 2005) of comet 9P/Tempel~1 before, during,
and after the Deep Impact event,
we used the Supernova Integral Field Spectrograph (SNIFS) 
(Aldering et al. 2002) to
obtain spectral data cubes covering the wavelength range from
350~nm to 1000~nm. This data acquisition method ensured that
spatial, spectral, and temporal information was obtained over
a wide wavelength range, without having to pre-select specific
filter bandpasses or slit orientations before the event. It was therefore
very well suited for an event where a poorly understood range of phenomena was
anticipated.
We are also presenting data from the VNIRIS instrument at the Lick Observatory
3 m telescope. This instrument is a more conventional long-slit spectrograph
and served to check our data at the limits of SNIFS' wavelength coverage.
VNIRIS is a three-channel instrument with a wavelength coverage from
500~nm to 2200~nm and extended our wavelength coverage into the infrared.

After the impact event, it became clear that the most interesting
aspects of our data were the spatial distribution and
temporal evolution of the CN~(0-0) emission at 388~nm and of the [OI] emission at 630~nm
as well as the evolution of the continuum scattered light color
in the hours after impact.

We will briefly describe the SNIFS instrument and the 
observing procedures and conditions in chapter 2.
The data reduction steps used to extract specific
information from the data cubes will be covered in 
chapter 3. Chapter 4 will discuss the results.

\section{Observations}

\subsection{The Supernova Integral Field Spectrograph SNIFS}
Comet 9P/Tempel~1 was observed in the nights of July 2 - 9, 2005 UT at the
UH 2.2m telescope using the SNIFS spectrograph
that is permanently mounted at the bent Cassegrain focus. The instrument
is described in more detail by Aldering et al. (2002) and Lantz et al. (2004). 
The SNIFS instrument is designed to
obtain photometrically calibrated spectra of supernovae against the
background of their host galaxy. It is therefore well suited to
obtain spectrophotometry of the gas and dust released from the comet nucleus by the impact
of the Deep Impact probe against the 
background of the more extended coma. 

Among the instruments used in the Earth-based Deep Impact observing campaign
(Meech et al. 2005), SNIFS is unique in its capability to obtain photometrically
calibrated spectral and spatial data over a very wide optical wavelength range.

SNIFS basically consists of a blue and a red spectrograph arm. After the two 
wavelength ranges are separated by a dichroic beamsplitter, the light in each
arm is focused
on a separate 15$\times$15 lenslet array with 0.4$\arcsec$ lenslets, forming
a contiguous 6$\arcsec$$\times$6$\arcsec$ field of view.
Each lenslet produces an image
of the telescope pupil, which then becomes effectively the entrance point
into a grism spectrograph. Each of the 15$\times$15 pupil images is dispersed and
produces a spectrum on the CCD detector. The lenslet array is rotated at such
an angle around the optical axis 
and against the orientation of the lenslet array
that the 225 individual spectra do not overlap. 
The dispersion is 0.22~nm/pixel in
the blue channel, the resolution is about 2 pixels, giving a 
resolving power of $\lambda/\Delta\lambda\approx$1000 at 440~nm. 
The red channel has a dispersion of 0.29~nm/pixel and a resolving power of
$\lambda/\Delta\lambda\approx$1300 at 760~nm (Lantz et al., 2004).
Spatially, one pixel of the extracted spectrum corresponds to one 0.4$\arcsec\times$0.4$\arcsec$
lenslet element. The spectral resolution is not dependent on the image quality in
the telescope focal plane.

\subsection{Observing Procedures and Conditions}
Data were obtained on the nights of July 2 and 3 (UTC) prior to impact,
on the night of July 4 (the impact night), when the first useful frame was exposed during
the moment of impact, and on the nights of July 5, 7 and 8, after the impact.
The nights of July 6 and 9 were lost due to a combination
of poor seeing
($\approx$2$\arcsec$) and cirrus clouds. 
The integration time used for comet spectra was 300~s on July 2 and 3, 90~s on July 4 and 5,
and 180~s on July 7 and 8.
The seeing in these nights was, of course, dependent on time, wavelength and air-mass.
Typically, the FWHM of standard star frames was $\approx$0.7$\arcsec$ in the red
channel of SNIFS and $\approx$0.85$\arcsec$ in the blue channel.
The Deep Impact event was observed at a geocentric distance of 0.89 AU, resulting
in a scale 
of 645 km arcsec$^{-1}$. The heliocentric distance was 1.51 AU.

Up to the night of July 7, the telescope was tracking the non-sidereal motion
of comet Tempel 1 in open loop, without guiding. Keeping the comet nucleus
centered in the field of view under these conditions proved rather difficult. We selected only
those frames for analysis where the full photometric aperture was within
the field of view, to avoid any possible color effects from extrapolating
from smaller apertures. 
However, we were using frames where the photometric
sky annulus was not fully contained in the field, a situation that the
IRAF apphot package is designed to handle. 

Observations of the comet were interdispersed with calibration observations,
so that our coverage of the post-impact phase was not continuous.
Observations of a position 5$\arcmin$ away from the comet nucleus, outside of 
the comet's coma, were taken as ``empty sky'' frames and later used in the
analysis of some of the data. Following the established procedures
for absolute photometric calibration of SNIFS data, solar analog stars
with known absolute flux calibration
were observed several times per night. Internal calibration data, a flat field
exposure and a spectral lamp exposure, were also taken at every telescope position
used for observations on the sky.

The night of July 4, 2005, UT, was photometric on Mauna Kea. The CFHT sky-probe
data in the first half of that night were all taken in the vicinity of comet Tempel 1, since CFHT was also
observing the Deep Impact event.
We did not have the time during the impact event to determine the extinction
independently.
The nights of July 2, 5, 7, and 8 (UTC) were also photometric. The night of July 3
was not fully photometric, but usable. 

\subsection{Lick Observatory VNIRIS Observations}
We are also presenting a spectrum obtained at the 3m telescope of Lick Observatory
with the Visible and Near-Infrared Imaging Spectrograph (VNIRIS) to double-check our
main conclusions and guard against artifacts introduced by the use of the innovative
SNIFS instrument. The VNIRIS is a more conventional spectrograph than SNIFS. It
has three long slit spectrograph channels, one optical channel, using a deep-depletion
CCD, and two infrared channels. For the observations reported here, a slit width of
2.7$\arcsec$ and nodding by 40$\arcsec$ along the slit
was used. The slit orientation was parallel to the parallactic angle to put atmospheric
refraction along the slit. The integration on the post-impact ejecta cloud was centered
on 6:42 UT ($\approx$ 50 min post-impact) with a total integration time of 12 minutes, 
the air mass for the comet observations
from Lick Observatory was about 3. A standard star (HD 126053) of
spectral type G1V was observed as a solar analog star, but unfortunately not at exactly the
same air mass.

\section{Data Reduction}
\subsection{Spectral Datacube Extraction}
The initial steps of the data reduction were done by the
SNIFS data reduction pipeline 
(Aldering et al. 2002, Aldering et al. 2006).
These steps were to remove
instrumental artifacts from the CCD frames and to extract
wavelength calibrated, extinction corrected and flux calibrated
spectra for each of the 15$\times$15 lenslets of the integral field units.
Using bias frames, dark frames, continuum flatfields, and an arc lamp spectrum
(He/Hg/Cd for the blue arm and Ne/Ar/Xe for the red arm),
the CCD data were bias and dark-current corrected, were flatfielded, cosmic rays were removed, 
and the position of each lenslet spectrum and its wavelength calibration on the
CCD frame was determined. 

To calibrate the flexure effects in
the instrument that are telescope-position dependent, at least one arc lamp frame
was taken at essentially the same telescope position where a sequence of science data was
taken. In the interest of observing efficiency during the impact time period, however,
we did not take an arc calibration frame prior to every single exposure.
As a result, the SNIFS real-time data reduction pipeline was run again after the completion of
all the observations of the night so that the association of calibration frames to
science frames could be optimized.

After extraction of each spectrum, the data were photometrically calibrated
using flux calibrations derived from the observations of solar analog stars,
including a standard extinction correction for clear photometric nights
on Mauna Kea.
The output of the SNIFS data reduction were
cubes (x,y,$\lambda$) stored as Euro-3D FITS tables
(Kissler-Patig et al. 2003), with x and y being the
position of each individual lenslet in the integral field units. 

\subsection{Extraction of Spectral Images}
From this state of the data reduction on,
SNIFS usually reduces its supernova observations with a data reduction pipeline
carefully tuned to the needs of the Nearby Supernova Factory project. While these
pipeline data products were very useful for an initial inspection of the
data from the Deep Impact observations, we decided for the final analysis of
the data to bypass these photometric extraction functions of the SNIFS pipeline and
instead to extract the photometric information directly from the Euro3D 
data cubes using IRAF (Tody, 1986) scripts.
The main reason was to achieve better control of atmospheric dispersion in
the extraction of broad-band spectrophotometry from the data cubes, better elimination
of residual cosmic ray events in the data, and a better handling of the changing 
flux distribution
over the course of the observations due to the expansion of the debris cloud around
the comet.

Initially, the 15$\times$15$\times\lambda$ data cubes were processed into 225$\times\lambda$
two-dimensional spectra. In this format, the few residual cosmic ray events that
had escaped automatic removal could be
easily identified and interactively eliminated using the IRAF image editing task (imedit).
As an example, Fig.~1 shows the extracted spectra in the blue and red channel in this two-dimensional
format, which is essentially a set of 15 long (6$\arcsec$) slit spectra stacked in
vertical direction. 
Note that for Fig.~1, we show a spectrum after subtraction of an
``empty sky'' frame taken a few minutes before the comet spectrum at a position 5$\arcmin$
from the comet nucleus, outside of the coma. The sky subtraction was done to eliminate
night sky emission lines from Fig.~1 which would otherwise limit the visibility of faint
emission features in the comet's coma. 
In this extracted, long-slit data format (Fig.~1), the CN (0-0) emission band at 388~nm
is seen prominently. The C$_3$ emission band at 405~nm is also faintly indicated in
the blue channel, but the signal-to-noise ratio is too low to analyze this emission in
any more detail. The C$_2$ Swan bands were not clearly detected. The brightest C$_2$ band at 
516~nm, in particular, lies in the overlap region of the blue and red
SNIFS channels, where throughput is low and the resulting signal-to-noise ratio is poor. 
In the red spectrum
the [OI] line at 630~nm is seen in the comet coma, and the red CN (1-0) band (911-932~nm)
is faintly indicated in Fig.~1. This figure is a sky-subtracted spectrum, so
telluric [OI] emission is removed to the degree that it is stable on timescales of a few minutes.
Also, the line seen at 630~nm in Fig.~1 is clearly stronger near the comet nucleus than at the
edge of the field of view, showing that it originates in the comet.
However, it should be noted
that for the extraction of spectrophotometric information by aperture photometry that we will
describe later, 
we did not subtract a separate sky frame,
since the sky subtraction was done as part of the aperture photometry procedure by measuring the
``sky annulus'' around the object on the object frame itself.

An extracted sky-subtracted spectrum of the comet 
near the time of maximum post-impact continuum flux (one half hour after impact) in a
2.4$\arcsec$ (6 pixel) diameter aperture is shown in Fig.~2. This spectrum is 
a reflectivity spectrum, obtained by dividing the comet spectrum by a spectrum
of a solar analog star. The blue and red channel of SNIFS were joined at 520~nm,
relying purely on the instrumental flux calibration without any further adjustments.
In Fig.~2, we also plot (as small symbols) data from both the blue and red arm of SNIFS that we do
not consider reliable. In both channels of the SNIFS spectrograph, the data points
at the short and long wavelength end of their respective range lie below the extrapolated
spectrum, and in the case of the overlap region around 520~nm, lie mostly below
the other channel. 
The seamless transition between the two spectrograph arms
illustrates the reliability of the instrumental calibration in that transition region. 
The spectrum shows
the violet emission band of CN, the C$_3$ band, and the emission line of [OI]. The feature at 763~nm
is an artifact from improper correction of the telluric O$_2$ absorption, the strongest
atmospheric absorption feature in the spectral range of our observations. 

For comparison, the extracted VNIRIS spectrum is shown in Fig.~3. The slit width
was 2.7$\arcsec$ and the extraction along the slit was limited to the profile of the
unresolved ejecta cloud at the time of the observations. Therefore, the spectrum in Fig.~3
is dominated by the unresolved ejecta cloud and the extended emission from the coma is
strongly suppressed.
Also, by comparison with the VNIRIS spectrum, it is clear that the
drop in signal at the longest wavelength end (between 950 and 1000~nm) in the SNIFS
spectrum is not real.
The VNIRIS spectrum shows a feature at $\approx$~920~nm from improper correction
of a telluric H$_2$0 absorption feature, but this is distinct from the feature seen in
the SNIFS data. This absorption feature is much more pronounced in the VNIRIS data due
to the high airmass of the observations, the difference in airmass between object and
standard, and the relatively low elevation of the Lick Observatory site.

Both for the extraction of emission band fluxes and for broad-band continuum photometry,
spectral images were extracted from the data cubes by integrating over the selected
wavelength range and by arranging the spatial dimensions into the proper orientation. 
Examples of these small 6$\arcsec\times$6$\arcsec$ images are shown in Figs. 4 and 5.

\subsection{Aperture Photometry}
Aperture photometry was obtained with the IRAF apphot package. 
The aperture photometry signal is the sum of the flux of
all pixels fully contained in the ``object aperture'', minus an estimate of the ``sky''
flux interpolated from the average flux in a surrounding ``sky annulus''. In the 
specific case of comet observations, the ``sky annulus'' also contains flux from the comet
coma, which consequently gets subtracted from the flux in the aperture. Therefore, our aperture
photometry is sensitive only to flux concentrated in the small photometry aperture, and not
to more extended flux. This characteristic of the aperture photometry leads to the sharp flux 
maximum and subsequent rapid decline of continuum and CN aperture photometry in Fig.~6.
The aperture
photometry software treats fractional pixels in the integration aperture properly,
a feature important for the small images and relatively coarsely sampled (0.4$\arcsec$) data obtained
by SNIFS. Also, apphot generates error messages when the integration aperture exceeds
the frame boundaries, but works in conditions where the sky annulus exceeds the frame
boundaries, not an infrequent occurrence on the small SNIFS frames.

\subsection{Emission Band Images}
\subsubsection{CN Emission}
The emission of the CN (0-0) band is the highest signal-to-noise emission feature in our data and
allows a detailed study of its temporal and spatial evolution.
Spectrophotometric information was extracted in a 10 pixel (2.9~nm) wide spectral window that included
all the flux from the CN band. The adjacent continuum was measured in two windows of the same
size, at the adjacent shorter and longer wavelengths. Since the solar scattered light
contains numerous absorption lines in this wavelength range, we divided the 2-dimensional
comet spectrum by the spectrum of solar analog star P041C, 
normalized to unity signal at the center
of the bandpass used to extract the CN emission
(Colina and Bohlin 1997, and Bohlin, Dickinson, and Calzetti 2001). 
The underlying continuum spectrum, normalized by the solar analog spectrum, was then estimated by the
average of the two adjacent continuum wavelength intervals and subtracted from the spectral
interval containing the CN emission. The resulting 
continuum-subtracted 
CN emission spectral frame was
averaged in wavelength and rearranged into a spatial image of the CN flux distribution.

Information on the CN flux was extracted from this continuum subtracted CN emission image
in two complementary ways: First, the average flux over the full 6$\arcsec\times$6$\arcsec$
field of view
was computed using the IRAF task ``imstat''. This gives, in essence, the surface brightness of
CN emission in the inner coma of comet Tempel 1, since the CN emission extends beyond the limits
of our field of view. The varying centering of the comet coma in our images injects some additional
noise even in these field-averaged data, but for times before and long after the impact, when
CN was much more extended than our 6$\arcsec\times$6$\arcsec$ field of view, the data are reliable.
Second, we obtained aperture photometry with an object aperture diameter of 7 pixels
(2.8$\arcsec$ = 1800 km) and sky annulus between 3.2$\arcsec$ and 4.8$\arcsec$ diameter. 
The CN photometry aperture was
centered on the flux in the 350~nm~-~400~nm integrated image, which was dominated by scattered continuum
light and had enough signal-to-noise for reliable centroid computation. 
This procedure avoided the problem that the ``apphot'' centering algorithm might otherwise peak up on random noise
spikes in the low signal-to-noise CN image. 
By using a continuum wavelength range that included the wavelength of the CN emission, effects
of differential atmospheric refraction on the position of the extraction aperture were avoided.
The 350~-~400~nm continuum images and a median-filtered version of the CN emission images are shown
in Fig.~4 and both the CN flux averages and aperture photometry are included in Fig.~6. Note that
the median filtering of the CN images was done only for clarity in Fig.~4. The photometry 
was extracted from the unfiltered CN images.

\subsubsection{[OI] Emission}
The [OI] emission at 630~nm was extracted over a wavelength range of 0.9~nm (3 pixels) centered on the line,
and the continuum was measured by interpolation between adjacent continuum bands, and then subtracted.
In the same way as for the CN emission, the comet spectra were divided by a normalized spectrum of
the solar analog star P041C prior to the extraction of the line emission. 
The 3.5~pixel radius (2.8$\arcsec$ diameter) photometric aperture
was centered on the centroid of the 600 - 650~nm continuum spectral image. 
The comparison of 600~-~650~nm continuum and the continuum-subtracted [OI] flux is shown in the
sequence of images in Fig.~5. 
The sky-annulus subtraction used as part of the aperture photometry eliminated any contribution
of telluric [OI] emission.
The aperture photometry of [OI] emission is included in Fig.~6 as filled circles. 

\subsection{Broad-band Aperture Photometry}

The spectral data cubes were integrated over the bandpasses listed in Table 1. 
The bandpasses cover the whole spectrum from 350~nm to 950~nm, with the exception of a gap
at $\lambda$~$\approx$~525~nm, in the beamsplitter transition region between the blue and red arm of
the SNIFS instrument. Also, the region of heavy telluric O$_2$ absorption at $\approx$~763~nm was avoided.
The data integrated over the bandpass were
rearranged as a set of 2-dimensional synthetic bandpass images. Aperture photometry using the
IRAF ``apphot'' package with an integrating aperture of 2.4$\arcsec$ (1550 km) diameter was then
performed. 
The SNIFS data are recorded in separate blue and red arms of the spectrograph.
The boresight of these two arms are not precisely identical. In addition, for observations at
higher airmass, atmospheric dispersion leads to a wavelength dependence of the 
centroid position, as is evident by a comparison of the images obtained in blue (Fig. 4)
and red (Fig. 5) bandpasses. Therefore, for the continuum photometry of comet 9P/Tempel~1, we
centered the photometry aperture individually on each bandpass image.

The data are presented in Figs. 7a and 7b as instrumental magnitudes, so that changes in brightness
lead to a linear shift of the spectral distribution. All data are normalized to
the flux distribution of the Sun. To do this normalization, a solar analog star, P041C,
was observed several times during the night of July 4, 2005, (UTC). As discussed by Colina and Bohlin (1997) and
Bohlin, Dickinson, and Calzetti (2001),
P041C has an effective temperature of 5900 K, slightly higher than that of the Sun (5777K).
The flux ratio of P041C to the Sun is given by Bohlin, Dickinson, and Calzetti (2001) 
and our measured spectrophotometry
was corrected accordingly. All data in Figs.~7a and 7b are normalized to this
corrected P041C flux by subtracting the corrected average instrumental magnitudes of P041C,
and therefore represent a normalization to the solar flux distribution.
In Fig. 7a and 7b , we show the solar analog standard stars observed on
three nights in the same normalization.
The data on P041C show a slightly blue spectrum in Figs.~7a and 7b,
while the spectral energy distribution of P177D shows small excesses both at short and
long wavelengths, both consistent with the results by Colina and Bohlin (1997). This good agreement in
the colors of the standard stars gives us confidence that the spectrophotometric calibration
of the SNIFS instrument is stable over the observing period.
The observed changes in the spectrophotometry of comet Tempel~1
in the hours after impact are therefore considered real.

The photometric errors were estimated from repeated observations of the comet on July 5, 2005 (UTC),
i.e., after the rapid changes in brightness induced by the impact had subsided.
The largest error comes from changes in the image quality due to seeing changes, focus 
drifts, and tracking errors and affect all wavelengths in nearly the same way, and possibly
also from intrinsic changes in brightness of the comet. For
the estimation of uncorrelated photometric errors, the photometric data were normalized to the same
magnitude average over the 625~-~825~nm range, where the signal to noise ratio was highest.
The rms scatter of the photometric values after this normalization of the average flux is 
represented by the error bars shown in Figs. 7a and 7b. 
These error bars are characteristic of the photometric errors when the comet was at or near
its pre-impact brightness, and also in the days after the impact when the nucleus and inner
coma had returned practically to pre-impact levels. While the comet was near its maximum
brightness post-impact, and even more so for the standard stars, the uncorrelated photometric
errors were smaller than the data point symbol in Figs. 7a and 7b.

\section{Discussion}

\subsection{Gas emission}
The formation and excitation of CN and [OI] emission is discussed
in detail in Manfroid et al. (2006) in this issue of Icarus.
The timescale for the creation of the CN radical from its grandparent and parent molecules under normal
steady-state conditions is long
compared with the duration of the July 4 post-impact observations.  Haser model
lifetimes for the parent of CN are of order 3$\times$10$^4$ s (Schleicher
and Farnham 2004).  The lifetime of the daughter against photodissociation
is 2$\times$10$^5$ s.  However, the typical lifetime is the
e-folding lifetime for dissociation.  Some amount of gas
is produced earlier, while other is produced later than this time.
Under normal, steady-state conditions, it is difficult to see gas with a timescale
different than the average.  However, the Deep Impact event was
very distinct from steady state so that the impulse in CN production 
can be seen in many data sets obtained of this event
(e.g. Cochran, Jackson, Meech and Glaz 2006).

On the other hand [OI] is a prompt emission, meaning that OI is directly
being produced in its excited states by photodissociation of one of its
many possible parent molecules (H$_2$, OH, CO$_2$, or CO). The $^1$D state responsible for
the 630~nm emission has a lifetime of $\approx$110 s. 
The 630~nm [OI] emission line is therefore spatially 
closely coupled to the location of the parent molecules and should
increase in proportion to the prevalence of the parent molecule.

\subsubsection{Temporal Evolution of CN Emission}

Fig. 4 shows a median filtered version of the input data for this CN photometry. In Fig. 6
the resulting CN aperture fluxes are shown as open triangles. 

Our last exposure of the first set of images (top row in Fig. 4)
ended at 06:07:25 (UTC) and did not yet show centrally concentrated CN emission. 
Our data first recorded centrally
concentrated CN emission in the exposure centered at 06:23:28 (UTC).
Observations at the Keck telescope by Cochran et al. (2006) show the first sign of CN emission in an exposure
lasting from 06:06:12 to 06:16:12 (UTC). 
These data are consistent if we assume that most of the CN flux in the Keck frame was
collected later in the exposure time, after 06:07:25 (UTC).
In combination, these data indicate that centrally concentrated CN emission became detectable
$\approx$ 20~minutes after the impact.
The CN lightcurve therefore trails the 350~-~400~nm integrated light curve (open stars in Fig.~6.)
that began to rise immediately after the impact.

In Fig.~4, we also show the CN images obtained in the minutes before impact, even though they
were not centered well on the comet.
While aperture photometry on these images was not possible and while the noise was elevated
due to the twilight sky brightness, we show them here for completeness.
The image taken at the time of impact (mid-exposure at 05:52:37 UTC)
shows a weak feature near, but not precisely
at the position of the comet.
Because of the discrepancy of the centroid positions, we conclude that the 
feature in this image is an artifact, and not a detection of the impact flash. Despite the position
discrepancy, the aperture photometry shows an elevated value for this image, which, by the same
argument, we also consider spurious, despite it formal 3$\sigma$ significance.

The signal average and standard deviation in the measurements in the night of July 7 and 8, after the rapid changes
have subsided  are 1.6 $\pm$ 173$\times$10$^{-20}$ Wm$^{-2}$, i.e. the CN aperture photometry signal is zero.
For the shorter 90s exposure times used on July 4, the noise would be expected to be 245$\times$10$^{-20}$ Wm$^{-2}$ rms.
Therefore, the detections of elevated flux levels in the hours after impact are significant at above
the 3$\sigma$ level. 

At the end of the observations on July 4, (UTC), about 3 hours after impact, the CN emission had spread
beyond the full field of view and the central concentration of CN emission had disappeared. Therefore,
the CN flux in the aperture and in the sky annulus were at the same level and the aperture photometry
signal consequently had returned to zero.
The average CN emission over the full 6$\arcsec\times$6$\arcsec$ field, however, peaked about 2 hours after impact, just at
the time when the impact-generated emission began to fill the field of view. It declined slowly
thereafter, as the impact-generated emission expanded beyond our field of view and flux therefore
got lost. 

The CN images in Fig. 4 have very low signal-to-noise ratio and do not allow a very precise measurement
of the CN expansion velocity. Under the reasonable assumption that during the middle set of images
in Fig. 4, roughly at 07:40 (UTC), $\approx$108 minutes after impact, the CN had reached the corner of the image
farthest from the comet (about 6$\arcsec$ $\approx$~3870~km), we get an average projected expansion velocity for CN and its parent
molecule of 0.6 km/s, in rough agreement with the measurement at Keck by Cochran et al. (2006).

On the night of July 5, 2005, the night following the impact, the average CN emission flux had stabilized to
a level slightly higher (12\%) than prior to impact, and remained at that level on the nights of July 7 and 8.
The average of all integrated CN flux measurements in the nights of July 2 and 3, in units of~10$^{-20}$ Wm$^{-2}$ in Fig.~6, and 
scaled to a 2.8$\arcsec$ aperture, is 2202, with a standard deviation of 92. The standard error of the average of 7 measurements
is therefore 35.
In the nights of July 5, 7, and 8, the average of 15 measurements is 2473 with a standard deviation of
69 and a standard error of the average of 18.
The change in CN flux average is therefore statistically significant. Since we used an instrument specifically designed
for spectrophotometry and since we used this instument continually through the Deep Impact observing
campaign under stable sky conditions, we are confident that we are not seeing systematic changes
in the instrument characteristics. We therefore believe that the 12\% change in CN flux integrated over
our field of view is real.
It is not clear whether the small increase in CN parent molecular release was caused by
the impact, or whether the change in CN production happened independently of the impact.
No concentration of the CN flux centered on the nucleus was found on the nights prior to and after the
impact night.
We did not find any indication for periodic variations in the integrated CN flux with the rotation 
of the comet nucleus. Such variations were
claimed by Jehin et al. (2006). None of our observations 2 day and 1 day prior to impact, and 1 day, 
3 days, and 4 days after impact would have included the maximum phase of their reported periodic
variations occurring 1.6 days prior and 0.1 day, 1.8 days, and 3.5 days after impact. Therefore, our
data can neither support nor contradict their finding.

\subsubsection{Temporal Evolution of [OI] Emission}

The [OI] flux follows the lightcurve of the broad-band flux quite closely.
Prior to the impact and in the days long after the impact, the [OI] flux concentrated
on the comet nucleus is detectable in our aperture photometry. The standard deviation
of individual photometric measurements on July 5 is 96$\times$10$^{-20}$ Wm$^{-2}$ at an
integration time of 90~s. The uncertainty of individual measurements on 
July 4, where the signal changed rapidly, should
be the same. Therefore, the detection and the changes detected are significant above the
5~$\sigma$ level. Post impact, on July 5, individual data points detect [OI] flux in
the aperture at the 3 $\sigma$ level. Assuming that the flux in this night was constant
over timescales of about half hour so that the scatter of the measurements represents
noise, the standard error of the average flux in this
night is a 6~$\sigma$ detection. Due to the longer integration time of 180~s, the
standard deviation in the nights of July 7 and 8 is only $\approx$ 60, leading to
a 6~$\sigma$ level of confidence in the flux averages in those nights.

In the first hour after impact, the [OI] flux rises by about a factor of 3, and then
slowly declines after peak flux in the aperture was reached 45 minutes after impact.
The decline in flux is slower than that of CN flux and of blue broad-band continuum flux.
In the night following the impact [OI] flux in the photometry aperture is down to
less than half of the peak flux, but still appears somewhat elevated above the level seen before
impact, even though this statement is uncertain due to the poor signal-to-noise of the
data. Three and four nights after the impact, the [OI] flux in the aperture is indistinguishable
from that before the impact. Spatially, the [OI] flux expands in unison with the continuum,
and much slower than the CN flux.

CN is probably a daughter species of a yet unidentified parent molecule,
whereas [OI] is both a daughter and
a granddaughter of H$_2$O and can also be produced from other parents.
Dissociation of H$_2$O (and other parent molecules) produces OI in its excited state
of short lifetime and the [OI] emission therefore traces the presence
of the parent closely.  Additionally, the impact released a great deal
of large H$_2$O ice particles which had a slower outflow velocity than the gas
and subsequently fragmented, with some producing [OI].  This more complicated
combined production and close link to the presence of H$_2$O ice and gas 
can account for the different scale lengths of [OI]
when compared with CN and for the different temporal evolution of the [OI] flux.

\subsection{Continuum Spectrophotometry}

\subsubsection{Reflectance Spectrum}
Overall, the spectrophotometry of comet Tempel 1 both before and after the impact indicates a red color of the scattered
light, similar to that found in most comets. In the logarithmic presentation (magnitudes)
chosen in Fig. 7a and 7b for the spectrophotometry, as well as in the linear plot in Fig. 2 and 3, 
we see a steepening of the spectral energy distribution at the shortest
wavelengths. In many comets, this region is dominated by gas emission (CN,C$_3$,and C$_2$).
By avoiding the emission features, as was done in our aperture photometry, or in comets with intrinsically low outgassing, evidence
for a steeper slope of the reflectivity spectrum at short wavelengths is evident, 
for example in Jewitt \& Meech (1986).
Outside of the emission and absorption features, the reflectivity of the ejecta cloud
after the Deep Impact event in the SNIFS observations can be roughly described by
two linear fits. From 350~nm to $\approx$580~nm, the normalized reflectivity gradient as defined
by Jewitt \& Meech (1986) is $\approx$22.6\% per 100~nm. From 580~nm to 940~nm, it is $\approx$7.5\% per 100~nm.
The two lines corresponding to these fits are included in Fig.~2.
It is noteworthy that the break between these two slopes does not coincide with the
wavelength break between the blue and red channel of the SNIFS instrument at 520~nm,
and is therefore considered real. 
Similarly, the VNIRIS spectrum is steeper at the short wavelength end, even though the
break is less pronounced than in the SNIFS spectrum. We suggest that the
small differences between these two spectra reflect residual calibration problems.
The tend towards a flatter reflectivity spectrum continues into the infrared, the spectrum measured by VNIRIS is
almost flat between 1500~nm and 2200~nm.

\subsubsection{Changes in Scattering Properties after Impact}
All the spectrophotometric data obtained while the photometry aperture was dominated by
impact ejected material show that this material had a bluer reflectivity spectrum than the material
usually released by the comet. The dotted line in Fig.~7a and 7b fitted to all photometric data points is the
average flux distribution in the data set taken about one half hour after impact, when the
ejecta cloud was still nearly unresolved and at its maximum brightness. 
The average post-impact maximum-light spectral distribution is shifted to match
the average
flux from 625~nm to 875~nm at other times.
The deviations from this fits at times prior to the impact, and in the days after the impact when
the inner coma of comet Tempel 1 had essentially returned to its normal condition, show that the
flux at short wavelenths was lower at those times.
The 375~nm data point is about 0.2 mag lower than
during the peak post-impact brightness. This is particularly well seen in the 7/5, 7/7, and 7/8
data points in Fig.~7b. For the quiescent state of the comet, the long wavelength points lie
systematically above the fitted impact ejecta spectrum, the effect is small ($\approx$0.05 mag).
Over the full range from 375~nm to 925~nm, we observe that the impact generated ejecta were $\approx$0.25 mag bluer than the
quiescent comet coma. This corresponds to a $\approx$4\% per 100~nm change in the slope of the spectrum
averaged over the wavelength range from 375~nm to 925~nm.

This change in color can be explained by an unspecific combination
of two effects. First, the particle size distribution of the material ejected after the impact 
may contain a larger fraction of very small particles, much smaller than the wavelengths of
visible light that leads to a bluer color of the Raleigh scattered light. 
Second, it may point to large quantities of pure water ice crystals
that are known to have blue optical reflection spectra (Lucey \& Clark 1985).
Infrared observations of the post-impact material have also found indications that the size 
distribution of impact-ejected material contained more small particles than were released
by the comet outside of the impact event (Harker Woodward and Wooden 2005) and (Sugita et al. 2005).

In combination, the changes in the comet's color in the hours after impact indicate that
the material ejected by the impact contains smaller particles and more ice, and is therefore
probably more pristine than the material released from the surface of the comet under normal conditions.

\section{Conclusions}
Our spectrophotometric observations of comet 9P/Tempel 1 in the days prior,
during, and after the impact of the Deep Impact probe allow 
the following conclusions:

1. The impact ejecta contained the parent molecule for the CN radical. A
noticeable level of CN emission centered on the newly created ejecta cloud
was first found twenty minutes after impact, while the maximum CN flux in
our field of view was measured about 2 hours after impact.

2. In the night following the Deep Impact, the CN emission integrated 
over our field of view had returned to a level about 12\% higher than pre-impact.
The CN emission then remained constant at this level 3 and 4 days after
the impact.

3. Emission in the [OI] line at 630~nm increased in the first one hour
after impact, and then levelled off to a nearly constant value for the
next few hours. One day after the impact, [OI] emission had gone back
to the pre-impact levels.

4. The impact ejecta contained, or soon fragmented into, a dust particle
size distribution with a larger fraction of small particles than the 
material normally released by the comet. As a result, the spectrophotometry
during the impact shows the scattered light from the expanding ejecta cloud
with slightly bluer color than the normal coma of the comet.

5. The reflectivity spectrum of comet Tempel~1 before, soon after, 
and long after the impact shows a steeper slope 
at $\lambda <$~600~nm than at longer wavelengths. The reflectivity spectrum at infrared
wavelengths is essentially flat.

\textbf{Acknowledgments}
This project was in part supported the The Aerospace Corporation's
Independent Research and Development Program.

\clearpage
\begin{deluxetable}{ccc}
\tabletypesize{\scriptsize}
\tablecaption{Spectrophotometry Spectral Bins}
\tablewidth{0pt}
\tablehead{
\colhead{Bandpass} & \colhead{$\lambda_1$} & \colhead{$\lambda_2$}
\\
\colhead{} & \colhead{[nm]} & \colhead{[nm]}
}
\startdata
375 & 350 & 400 \\
425 & 400 & 450 \\
475 & 450 & 500 \\
575 & 550 & 600 \\
625 & 600 & 650 \\
675 & 650 & 700 \\
725 & 700 & 750 \\
815 & 780 & 850 \\
875 & 850 & 900 \\
925 & 900 & 950 \\
\enddata
\end{deluxetable}

\newpage
\noindent
\textbf{References}

\noindent
Aldering, G., Adam, G., Antilogus, P., Astier, P., Bacon, R.,
Bongard, S., Bonnaud, C., Copin, Y, Hardin, D., Henault, F.,
Howell, D. A., Lemonnier, J.-P., Levy, J.-M., Loken, S. C.,
Nugent, P. E., Pain, R., Pecontal, A., Pecontal, E., Permutter, S.,
Quimby, R. M., Schahmaneche, K., Smadja, G., Wood-Vasey, W. M. 2002.\\
Overview of the Nearby Supernova Factory.\\
Proc. of the SPIE, Volume 4836, 61-72.

\noindent
Aldering, G., Antilogus, P., Bailey, S., Baltay, C., Bauer, A., Blanc, N.,
Bongard, S., Copin, Y., Gangler, E., Gilles, S., Kessler, R., Kocevski, D.,
Lee, B. C., Loken, S., Nugent, P., Pain, R., Pecontal, E., Pereira, R.,
Perlmutter, S., Rabinowitz, D., Rigaudier, G., Scalzo, R., Smadja, G.,
Thomas, R. C., Wang, L., and Weaver, B. A. 2006.\\
Nearby Supernova Factory Observations of SN 2005gj: Another Type Ia
Supernova in a Massive Circumstellar Envelope.\\
ApJ, in press, available on astro-ph 0606499.

\noindent
Bohlin, R. C., Dickinson, M. E., and Calzetti, D. 2001.\\
Spectrophotometric Standards from the Far-Ultraviolet to the Near-Infrared:
STIS and NICMOS Fluxes.\\
The Astronomical Journal, 122, 2118-2128.

\noindent
Cochran, A. L., Jackson, W. M., Meech, K. J., Glaz, M. 2006.\\
Observations of Comet 9P/Tempel 1 with the Keck 1 HIRES Instrument
during Deep Impact.\\
Icarus, submitted, this issue.

\noindent
Colina, L., Bohlin, R. 1997.\\
Absolute flux distributions of solar analogs from the UV to the near-IR.\\
The Astronomical Journal 113, 1138-1144.

\noindent
Feldman, P. D., Cochran A. L. and Combi, M. R. "Spectroscopic Investigations
of Fragment Species in the Coma" in Comets II pp. 425-448
Comets II is edited by Festou, M. C., Keller. H. U. and Weaver H. A.
The University of Arizona Press, Tucson 2004

\noindent
Fink, U., Hicks, M. D. 1996.\\
A survey of 39 comets using CCD spectroscopy.\\
\apj 459, 729-743.

\noindent
Harker, D. E., Woodward, C. E., Wooden, D. H. 2005.\\
The dust grains from 9P/Tempel 1 before and after the
encounter with Deep Impact.\\
Science 310, 278-280.

\noindent
Jehin, E., Manfroid, J, Hutsem\'{e}kers, D, Cochran, A. L., Arpigny, C.,
Jackson, W. M., Rauer, H., Schulz, R., Zucconi, J.-M. 2006.\\
Astrophysical Journal, in press.

\noindent
Jewitt, D., Meech, K. J. 1986.\\
Cometary grain scattering versus wavelength, or, ``What color is comet dust?''\\
The Astrophysical Journal 310, 937-952.

\noindent
Kissler-Patig, M., Copin, Y., Ferruit, P., P\'{e}contal-Rousset, A., Roth, M. M. 2003\\
``The Euro3D data format: A common FITS data format for integral field spectrographs''\\
Astronomical Notes 325, 159-162.

\noindent
Lantz, B., Aldering, G., Antilogus, P., Bonnaud, C., Capoani, L.,
Castera, A., Copin, Y., Dubet, D., Gangler, E., H\'{e}nault, F.,
Lemonnier, J.-P., Pain, R., P\'{e}contal, A., P\'{e}contal, E., Smadja, G. 2004.\\
SNIFS: a wideband integral field spectrograph with microlens arrays.\\
Proc. of the SPIE, Volume 5249, 146.

\noindent
Lucey, P. G., Clark, R. N. 1985.\\
Spectral properties of water ice and contaminants\\
NATO ASI series, V. 156,
Proc. NATO Advanced Research Workshop on Ices in the Solar System\\
Reidel, Dordrecht, p. 155-168.\\
The Astrophysical Journal 310, 937-952.

\noindent
Manfroid, J., Hutsem\'{e}kers, Jehin, E., Cochran, A. L., Arpigny, C.,
Jackson, W. M., Meech, K. J., Schulz, R., Zucconi, J.-M. 2006,\\
The impact and rotational lightcurves of Comet 9P/Tempel 1.\\
Icarus, in press, astro-ph 0608611.

\noindent
Meech, K. J., and 208 colleagues 2005.\\
Deep Impact: Observations from a Worldwide Earth-Based Campaign.\\
Science 310, 265-269.

\noindent
Sugita, S., Ootsubo, T., Kadono, T., Honda, M., Sako, S., Miyata, T.,
Sakon, I., Yamashita, T., Kawakita, H., Fujiwara, H., Fujiyoshi, T.,
Takato, N., Fuse, T., Watanabe, J., Furusho, R., Hasegawa, S.,
Kasuga, T., Sekiguchi, T., Kinoshita, D, Meech, K. J., Wooden, D. H.,
Ip, W. H., A'Hearn, M. F. 2005.\\
Subaru Telescope Observations of Deep Impact.\\
Science 310, 274-277.

\noindent
Tody, D. 1986.\\
``The IRAF Data Reduction and Analysis System'',\\
{\it SPIE Instrumentation in Astronomy VI}, Ed. D. L. Crawford, 627, 733-748.

\noindent
Schleicher, D. G. and Farnham, T. L. "Photometry and Imaging of the Coma
with Narrowband Flters" in Comets II pp. 449-470
Comets II is edited by Festou, M. C., Keller. H. U. and Weaver H. A.
The University of Arizona Press, Tucson 2004

\clearpage
\begin{figure}
Figure 1, Hodapp et al., Deep Impact Spectro-Photometry
\vspace{0.3in}
\figurenum{1}
\epsscale{1.0}
\plotone{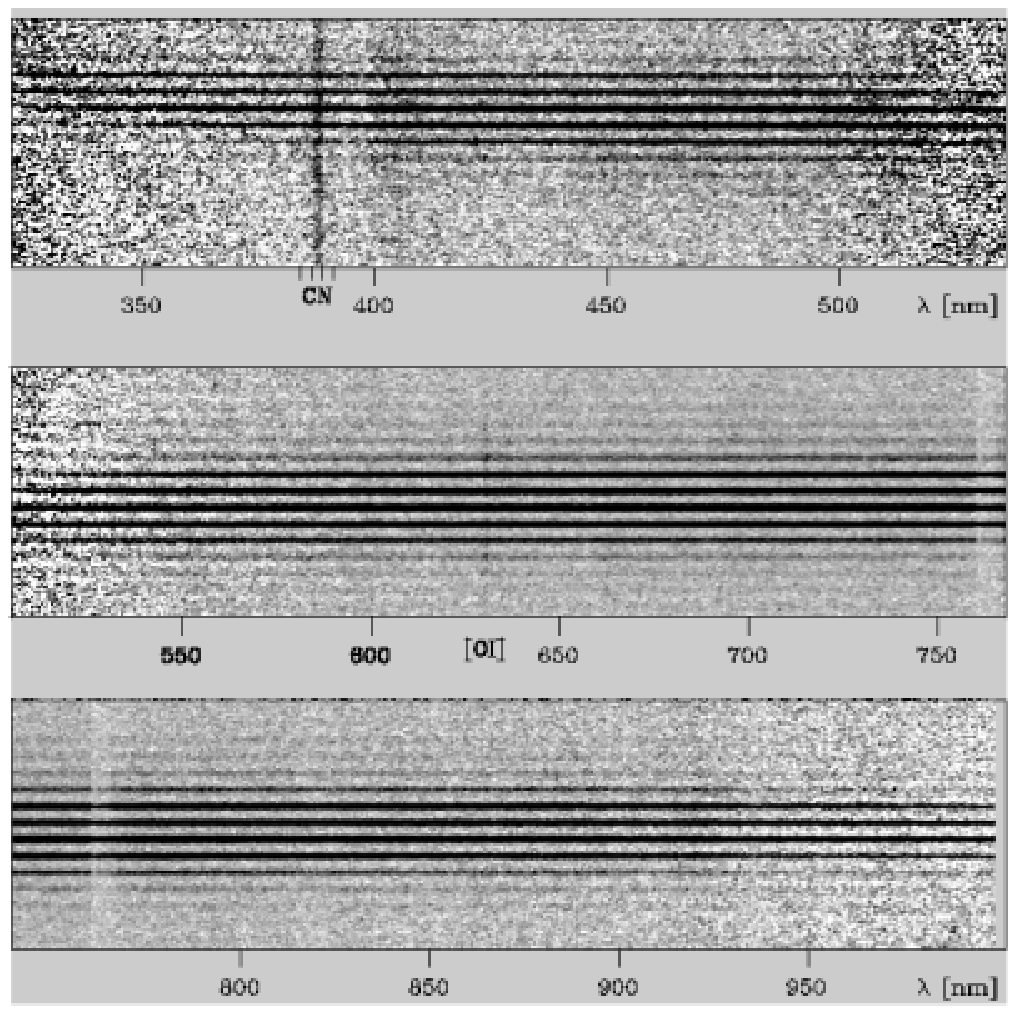}
\end{figure}
\nopagebreak

\clearpage
Figure 1.\\
Extracted SNIFS spectral datacubes in two-dimensional representation, where
the columns of the integral field unit are arranged along the vertical axis.
This presentation is essentially 15 long slit spectra, each 6$\arcsec$ long, stacked vertically.
The blue spectrum is on top, the red spectrum is the center and bottom panels. The dispersion
is different in blue and red spectra. 
The datacubes displayed here are the difference of a comet spectrum and a
sky spectrum taken 5$\arcmin$ from the comet. Therefore, no night sky
lines are visible. 
The violet emission of CN (0-0) is strongly visible. We are also faintly detecting the C$_2$ emission
band at 405~nm and the [OI] emission at 630~nm. The C$_2$ emission is too broad and faint
for further analysis.

\clearpage
\begin{figure}
Figure 2, Hodapp et al., Deep Impact Spectro-Photometry
\vspace{0.3in}
\figurenum{2}
\epsscale{0.8}
\plotone{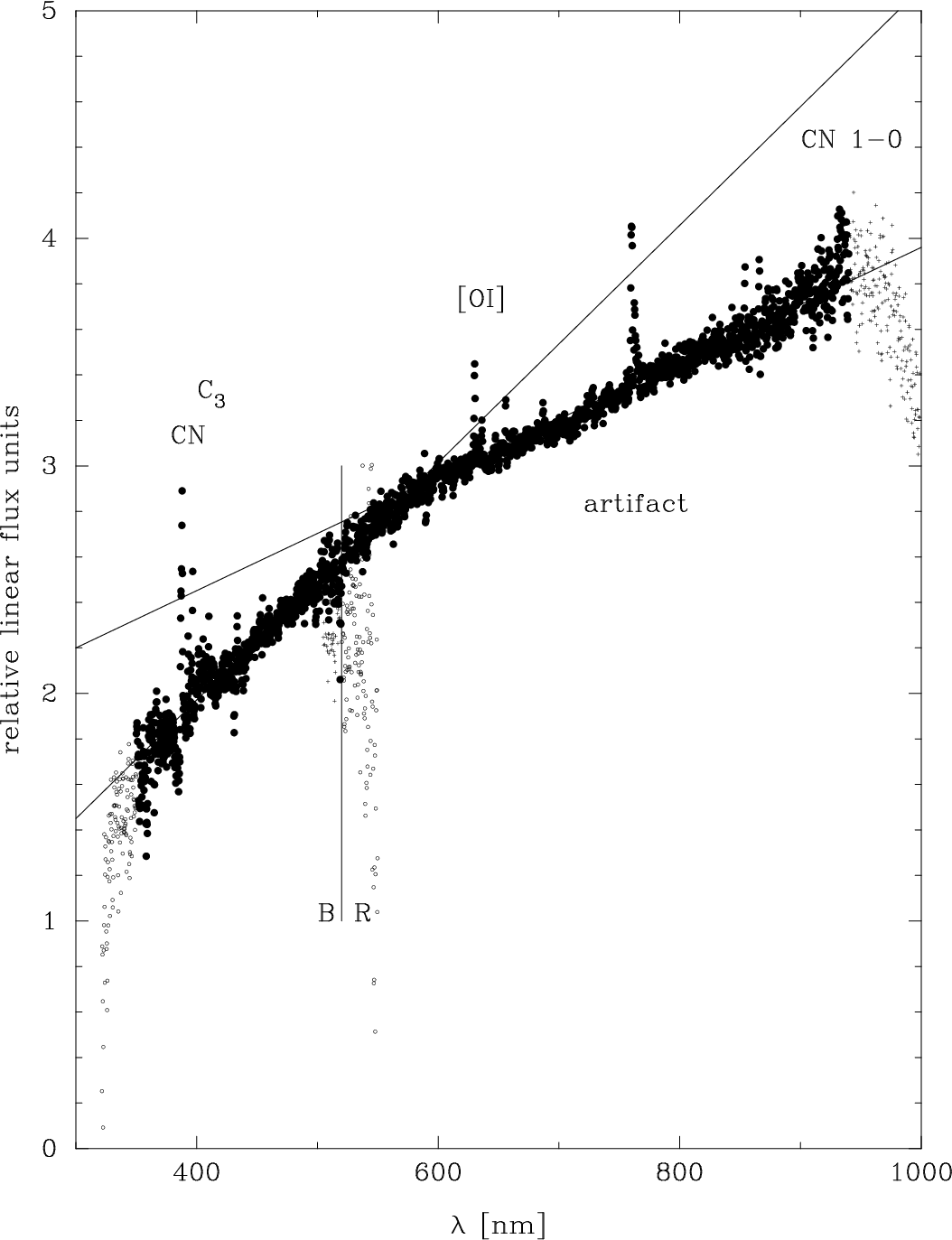}
\end{figure}
\nopagebreak

\clearpage
Figure 2.\\
Reflectivity spectrum of comet 9P/Tempel~1 after the impact, near the
time of maximum brightness in a 2.4$\arcsec$ aperture. The spectrum
is the ratio of the comet spectrum and that of solar analog star P041C
and is displayed here in arbitrary linear flux units.
The blue and red channel of the instrument were individually photometrically
calibrated and the data match very well without further adjustment.
The emission lines of CN and [OI] are clearly visible, and the C$_3$ band
is faintly indicated. The feature at 763~nm is an artifact from the strong
telluric O$_2$ absorption.
Filled circles represent the data that are considered reliable. We also
show the data at the ends of the spectral ranges of the blue spectrograph arm
(small open circles) and of the red spectrograph arm (small plus signs) to
illustrate that at the ends of each spectrum, the calibration was not reliable
and why these data have not been included in the analysis. The boundary between
where the blue vs. red data were used is at 520 nm and this boundary is indicated
in the Figure. This boundary does not coincide with the change in slope of the
spectrum. Below 580~nm, the normalized slope between 350~nm and 580~nm is 22.6\%,
between 580~nm and 940~nm, the normalized slope is 7.5\%.

\clearpage
\begin{figure}
Figure 3, Hodapp et al., Deep Impact Spectro-Photometry
\vspace{0.3in}
\figurenum{3}
\epsscale{0.8}
\plotone{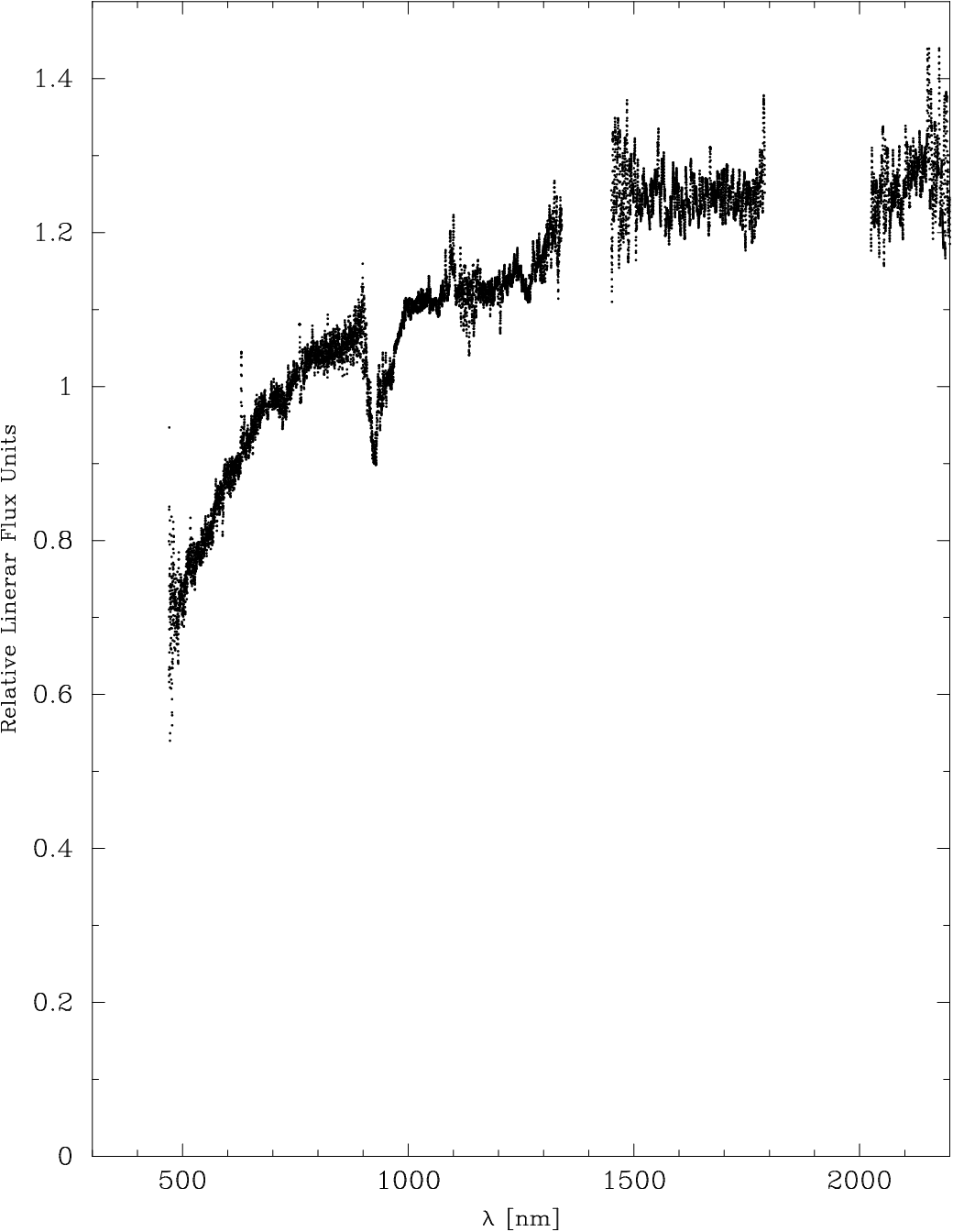}
\end{figure}
\nopagebreak

\clearpage
Figure 3.\\
Reflectivity spectrum in relative flux units, obtained with the
3-channel VNIRIS spectrograph at the Lick Observatory 3~m telescope. The spectrum
covers the wavelength range from 500~nm to 2200~nm. 
The integration time was 12 min, centered at 6:42 UT, 50 min after
the impact at the time of maximum brightness of the impact generated
ejecta cloud.
While this spectrum shows a poorly corrected atmospheric absorption feature at 920~nm
this feature is clearly distinct from the 950 - 1000~nm feature seen in Fig. 2.
Therefore, both features are artifacts.
The spectrum confirms the steeper reflectivity spectrum at shorter wavelengths compared
to the longer wavelengths.

\clearpage
\begin{figure}
Figure 4, Hodapp et al., Deep Impact Spectro-Photometry
\vspace{0.3in}
\figurenum{4}
\epsscale{0.6}
\plotone{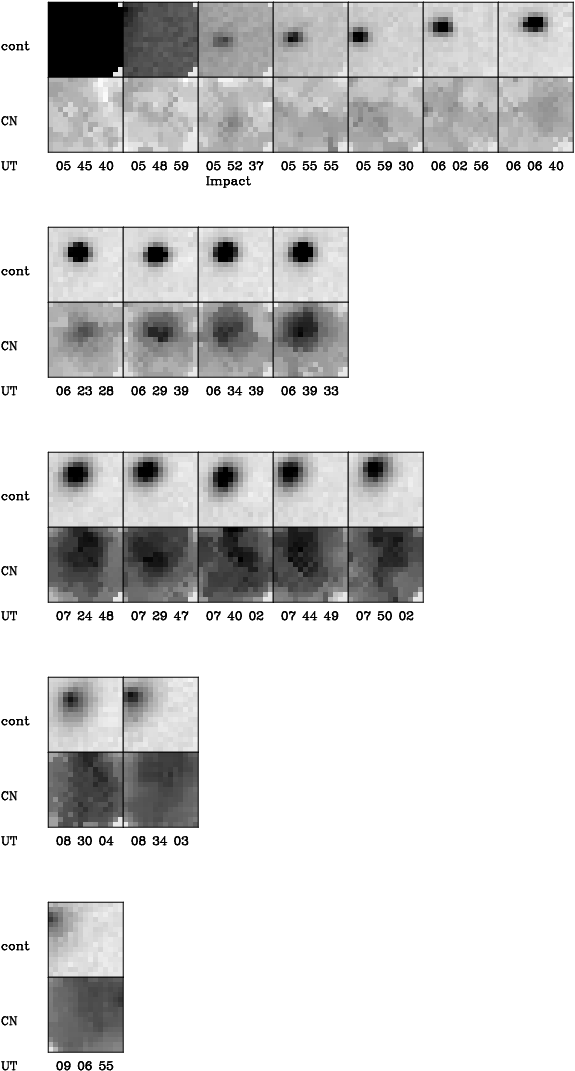}
\end{figure}
\nopagebreak

\clearpage
Figure 4.\\
The sequence of extracted, continuum-subtracted CN images during
the night of the Deep Impact event. The top two left frames were taken prior
to impact under twilight conditions, so the continuum image appears
very dark. In those two images, the comet nucleus was located in 
the upper left corner of the field of view. The third image was
exposed during the impact time. The weak peak found in this image
is considered spurious.
The gaps between groups of images indicate time when no comet data
were taken because of other observing tasks such as sky fields and
standard stars.
CN emission centered on the comet nucleus begins to be visible about
one half hour after impact, in the first frame of the second set of
data. The CN emission then rapidly increases in intensity and expands
spatially. About two hours after impact, the added CN emission fills the
whole field of view and the average flux levels are clearly higher than before
the impact. This is quantitatively displayed in the lightcurves in Fig.~6.

\clearpage
\begin{figure}
Figure 5, Hodapp et al., Deep Impact Spectro-Photometry
\vspace{0.3in}
\figurenum{5}
\epsscale{0.6}
\plotone{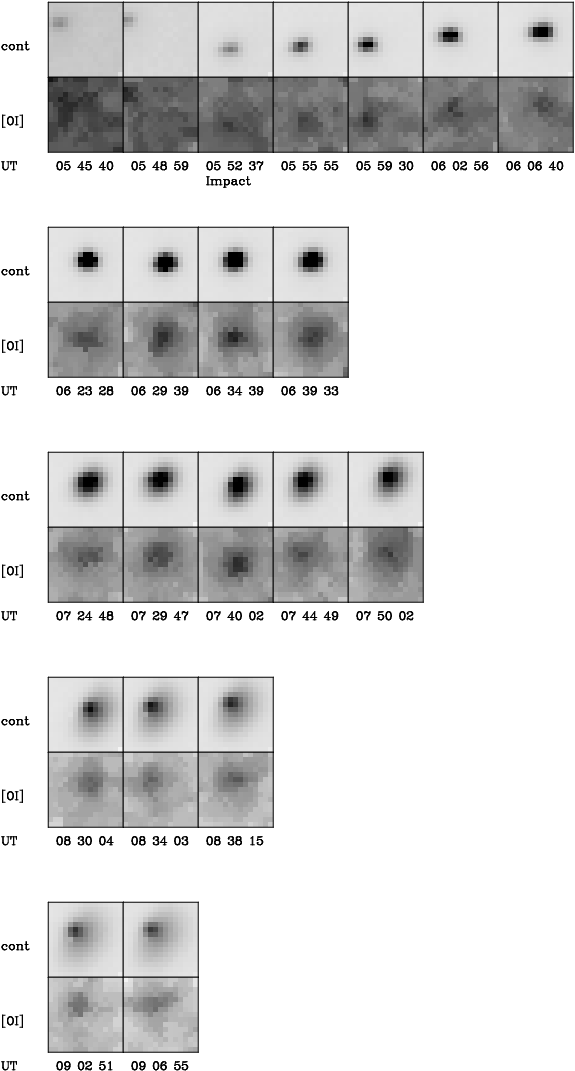}
\end{figure}
\nopagebreak

\clearpage
Figure 5.\\
The sequence of extracted, continuum-subtracted [OI] ($\lambda$~=~630~nm) images during
the night of the Deep Impact. 
In the first two images
the comet nucleus was located in 
the upper left corner of the field of view. The third image was
exposed during the impact time. 
The gaps between groups of images indicate time when no comet data
were taken because of other observing tasks such as sky fields and
standard stars.
There is a uniformly distributed component of [OI] emission from
night-sky emission. This component is subtracted out in the sky-subtracted
aperture photometry shown in Fig.~5. The 2.8$\arcsec$ diameter aperture
photometry shows a rapid rise of the [OI] emission centered on the comest
nucleus.

\clearpage
\begin{figure}
Figure 6, Hodapp et al., Deep Impact Spectro-Photometry
\vspace{0.3in}
\figurenum{6}
\epsscale{0.8}
\plotone{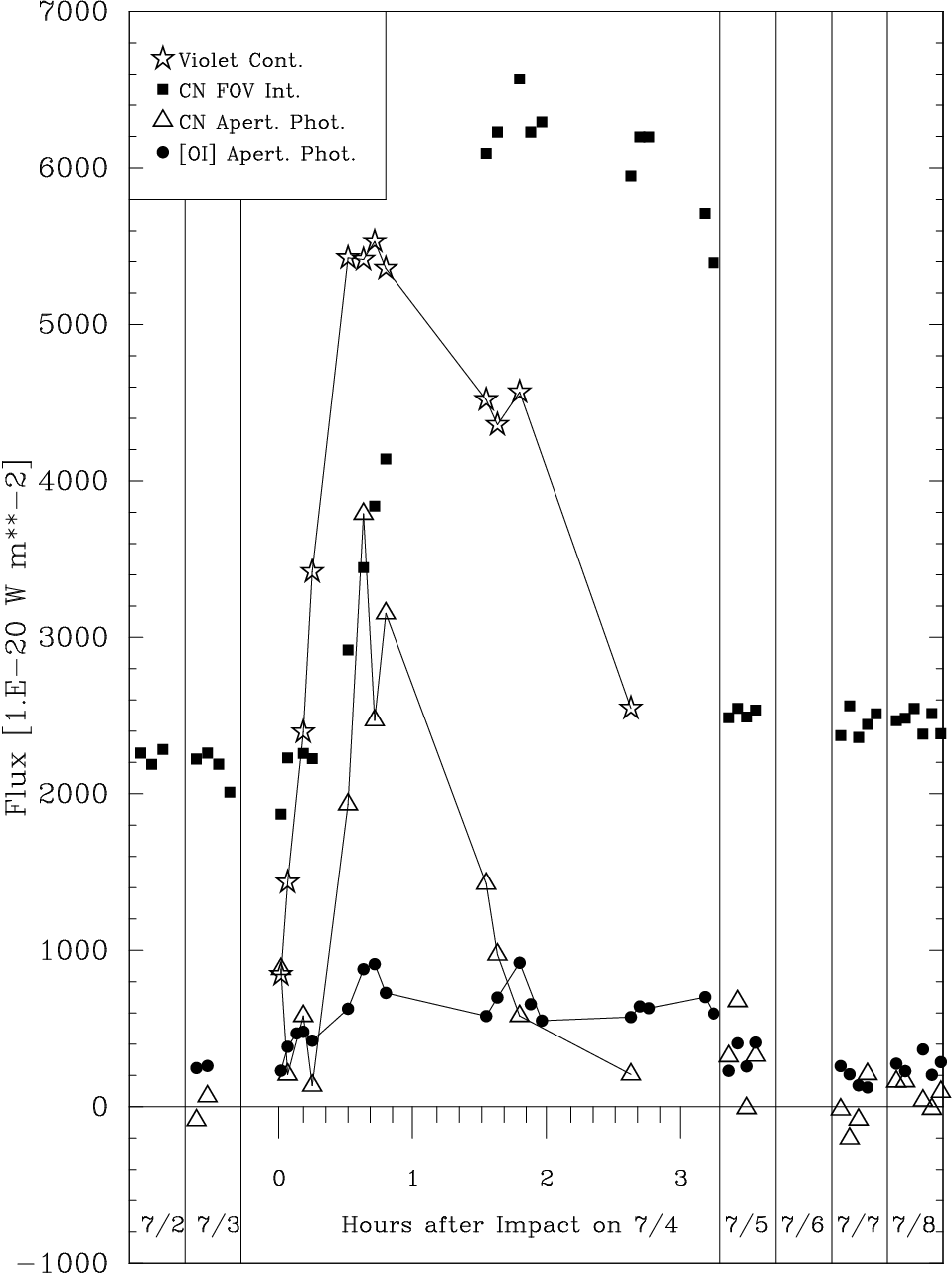}
\end{figure}
\nopagebreak

\clearpage
Figure 6.\\
This figure summarizes the photometry of comet 9P/Tempel 1
in broad bandpasses and in narrow bandpasses centered on the violet CN band and the [OI] emission
features.
The central panel of this figure are lightcurves obtained on July 4, 2005, UT, time is indicated
in hours after the time of impact of the Deep Impact probe. The narrower
panels to the right and left show individual measurements taken in the nights prior 
(July 2 and 3, UT) and after the impact (July 5,7,and 8). No data were obtained on July 6 due
to bad weather.
These data are shown in the
correct sequence, but without a specific time scale on the horizontal axis, since we did not
expect, nor observe, changes on timescales of less than one hour in those nights. 

Several measurements are plotted here:
Open triangles are aperture photometry on CN-images. 
Solid squares are the average CN-fluxes over the full field of 6$\arcsec$$\times$6$\arcsec$
corresponding to 3870~km~$\times$~3870~km at the comet.
The average fluxes over the field of view
(solid squares).
have been scaled to an area corresponding to the 2.8$\arcsec$ diameter aperture
used for the gas emission aperture photometry, to help with a direct comparison of the fluxes.
Filled circles are aperture photometry on [OI] images.
Open stars are 350-400~nm integrated images dominated
by scattered continuum, measured in 2.4$\arcsec$ apertures; 
these measurements have been scaled by a factor of 1/50 
for comparison with the line emission fluxes.
The signal is given in units of 10$^{-20}$ Wm$^{-2}$, based on the SNIFS instrument
absolute calibration and standard Mauna Kea extinction. We estimate the systematic
uncertainty of this value to be about 10\%.

\clearpage
\begin{figure}
Figure 7a, Hodapp et al., Deep Impact Spectro-Photometry
\vspace{0.3in}
\figurenum{7a}
\epsscale{0.8}
\plotone{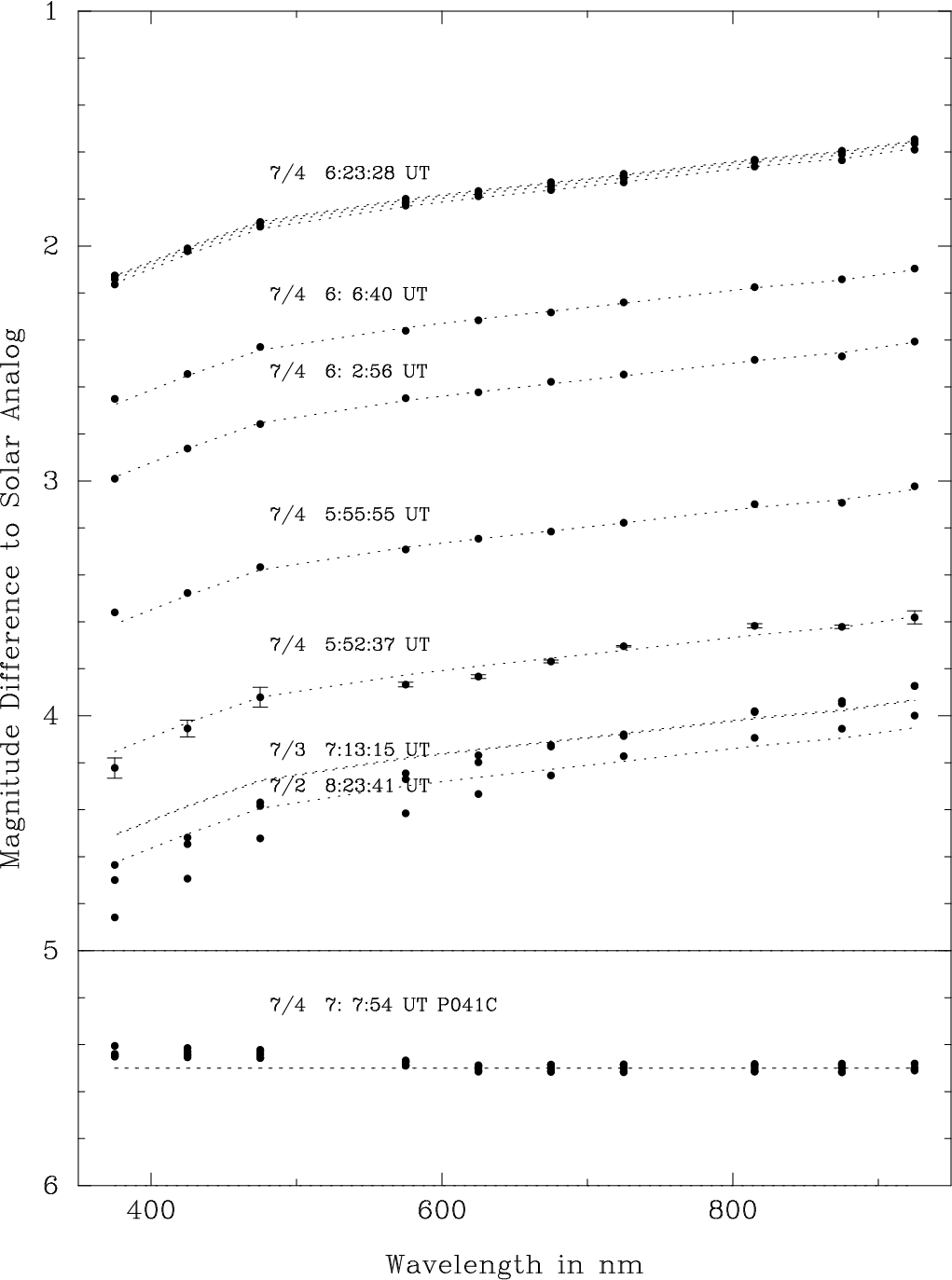}
\end{figure}

\clearpage
Figure 7 a.\\
The synthetic bandpass photometric measurements taken at different times are plotted against
wavelength. All data are normalized to a solar spectrum by subtracting the data
of solar analog star P041C and the appropriate flux correction.
The lower panel shows the measurements of P041C on 7/4.
The deviations from a constant signal are consistent
with the flux corrections for these stars. 
In the middle panel, the comet measurements for 7/2 (shifted down for clarity) and then for 7/3 
are presented. Then we plot the data taken in the first 
hour after impact on 7/4 without additional shifts so that the magnitude differentials
represent the true lightcurve of the impact event.. Even the first frame on 7/4
(taken during impact) is brighter than 7/3. In the 2.4$\arcsec$ (1550~km) aperture used
here for photometry, the brightness reaches a plateau about one half hour after impact.
These measurements about one half hour after impact are the highest signal-to-noise data obtained on the comet and
represent the state of the comet dominated by impact-generated ejecta. The average
of these measurements is the ``post-impact-maximum'' flux distribution (dotted line) that is fitted to the
other comet measurements. 
The measurements prior to impact show a redder continuum than the data dominated by impact
generated ejecta.
In other words, the impact-generated material had a bluer reflectivity spectrum
than the material released by the comet under normal conditions.

\clearpage
\begin{figure}
Figure 7b, Hodapp et al., Deep Impact Spectro-Photometry
\vspace{0.3in}
\figurenum{7b}
\epsscale{0.8}
\plotone{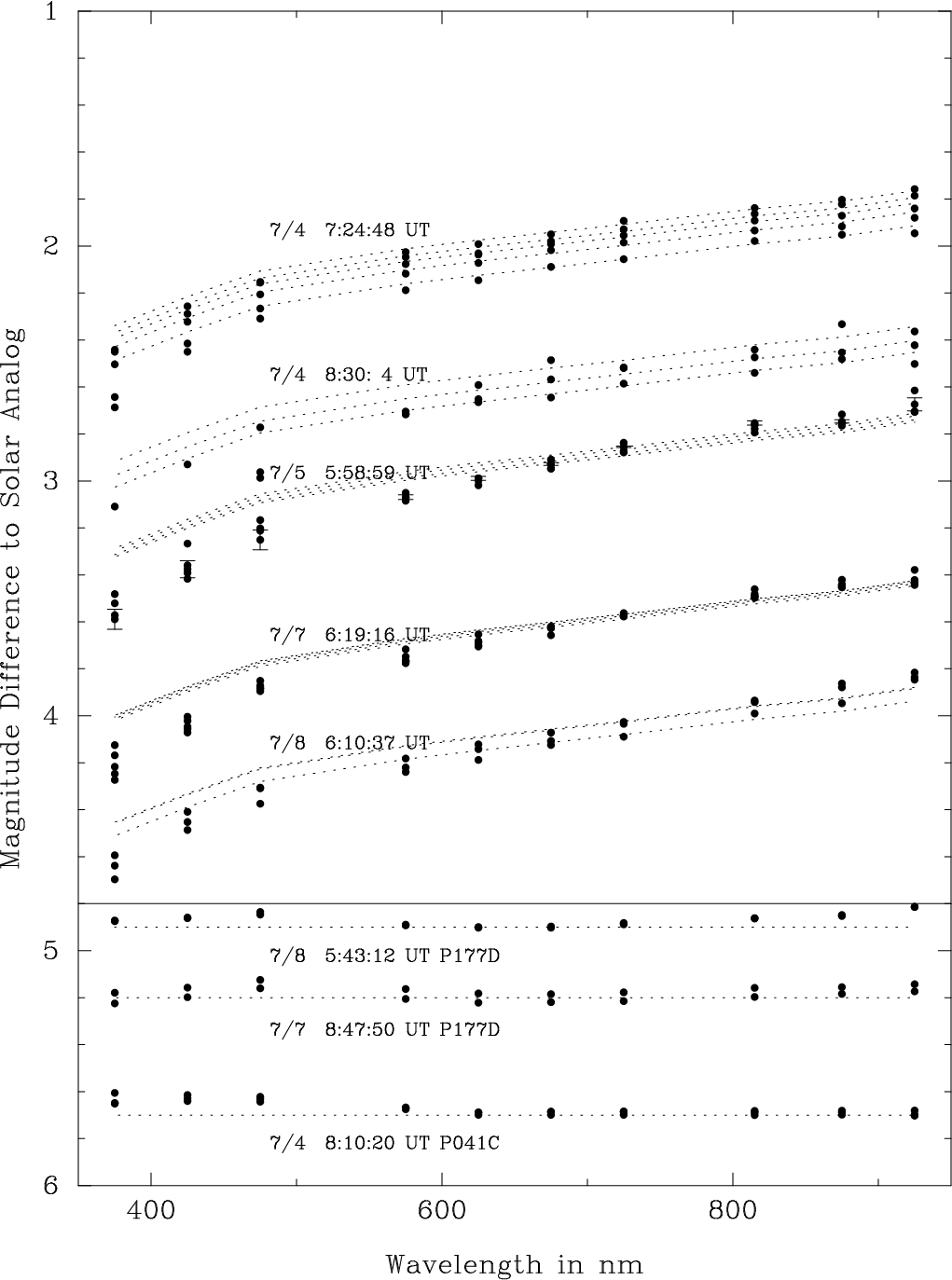}
\end{figure}

\clearpage
Figure 7 b.\\
The synthetic bandpass photometric measurements taken at different times are plotted against
wavelength. All data are normalized to a solar spectrum by subtracting the data
of solar analog star P041C and the appropriate flux correction.
The lower panel shows the measurements of P041C on 7/4 and of another solar analog
star, P177D, on 7/7 and 7/8. The deviations from a constant signal are consistent
with the flux corrections for these stars. 
The top panel shows the comet measurements (shifted for clarity) after the lightcurve
maximum on 7/4 and in the nights following
the impact. 
After maximum brightness, the color of the comet returned to the redder continuum that
was observed prior to the impact.

\end{document}